**Sub-nm Curvature Unlocks Quantum Flexoelectricity in Graphene**

Sathvik Ajay Iyengar[1], James G. McHugh[2], Jonathan P. Salvage[3], Robert Vajtai[1], Alan Dalton[4]*, Manoj Tripathi[4, 5]*, Pulickel M. Ajayan[1]*, Vincent Meunier[6]*

[1] *Department of Materials Science and NanoEngineering, Rice University; Houston, TX, 77005 USA*

[2] *National Graphene Institute, University of Manchester, Booth St. E., Manchester, M13 9PL, United Kingdom*

[3] *School of Pharmacy and Biomolecular Science, University of Brighton, Brighton BN2 4GJ, U.K*

[4] *Department of Physics and Astronomy, School of Mathematical and Physical Sciences, University of Sussex; Brighton BN1 9QH, United Kingdom*

[5] *Department of Civil and Environmental Engineering, South Dakota School of Mines and Technology, Rapid City, SD 57701*

[6] *Department of Engineering Science and Mechanics, Pennsylvania State University; University Park, PA, 16802, USA*

*Corresponding author. Email: Alan Dalton A.B.Dalton@sussex.ac.uk, M.Tripathi@sussex.ac.uk, ajayan@rice.edu, vincent.meunier@psu.edu

**Abstract:**

Flexoelectricity—polarization induced by strain gradients—is especially pronounced in two-dimensional (2D) materials due to their mechanical flexibility and sensitivity to mechanical deformation. In nanostructures with sub-nm curvature, this effect is governed by quantum-mechanical polarization and electrostatic modulation, not merely classical lattice distortion. Here, we present the first direct experimental and theoretical demonstration of large intrinsic quantum flexoelectricity in graphene nanowrinkles, exhibiting polarization densities ($P_{th} \sim 4$ C/m$^2$; $P_{exp} \sim 1$ C/m$^2$) that exceed those of mesoscale systems by 5 to 7 orders of magnitude. These nanowrinkles, with sub-nm radii of curvature at their apex, undergo atomic-level buckling and result in localized strain fields, as confirmed by sub-micron Raman spectroscopy. These curvatures create an asymmetry to π-orbital interactions across the atomic layer, which, in turn, leads to localized polarization densities. Kelvin probe force microscopy reveals curvature-dependent work function shifts consistent with flexoelectric polarization, while conductive atomic force microscopy detects reproducible flexoelectric currents exhibiting a threshold voltage ($\Phi_{th} \sim 1$ V) that matches the band offset predicted by ab initio calculations (~1.2 V). Together, these results confirm how flexoelectric dipoles reshape the local electronic potential. Graphene nanowrinkles thus provide a pristine platform for uncovering quantum-mechanical flexoelectricity—a fundamentally ubiquitous effect, whose study in the simplest crystalline material can illuminate electromechanical behavior across condensed matter, soft matter, and biological systems.

## Introduction

Flexoelectricity—the coupling between strain gradients and polarization—can occur in all materials, even those with centrosymmetry, and has long been recognized as a universal electromechanical phenomenon [1–3]. In low-dimensional nanomaterials, where extreme compliance allows for localized curvature and large strain gradients, the magnitude and consequences of flexoelectricity are predicted to be particularly pronounced[4,5]. Classical flexoelectricity occurs when charges are separated across a deformed membrane, leading to a net polarization. It can also arise from the asymmetric redistribution of dipoles upon bending. In contrast, quantum flexoelectricity results from changes in electronic orbital overlap and is thus predicted even in structures without pre-existing dipoles or charge separation. Kalinin and Meunier[6] first theorized that quantum polarization could be induced in otherwise non-polar van der Waals (vdW) atomically-thin, 2D crystals such as graphene. Since then, theoretical studies have expanded on these predictions in other xenes, 2D boron nitride, and transition metal dichalcogenides (TMDs)[7–9], yet direct experimental confirmation remains limited.

In contrast to bulk materials, where competing electromechanical phenomena obscure flexoelectricity[10], 2D systems offer a clean platform for directly investigating strain gradient-induced polarization at the nanoscale. While recent experimental efforts have begun to probe this research space, isolating intrinsic flexoelectric behavior remains challenging. For instance, flexo-photovoltaic responses have been reported in $MoS_2$-based hybrid heterostructures incorporating $VO_2$ phase-change layers[11], where strain gradients generate polarization-driven currents. However, the involvement of thermally activated transitions and structural complexity can pose challenges for reproducibility and scalability.

It is also important to note that access to flexoelectricity in 2D materials (via large-area thin-film bending in bubbles[12]) does not immediately imply a quantum manifestation—the prerequisite is that the buckling, deformation, or strain must be set by the spatial extent of the out-of-plane orbitals (such as π-orbitals in graphene), representing the thinnest possible quantum mechanical separation over which flexoelectric polarization can manifest. To that effect, most prior studies have accessed strain gradients corresponding to ~$10^6$ $m^{-1}$ curvature[13,14], several orders of magnitude below the ~$10^{10}$ $m^{-1}$ gradients achieved here through atomically sharp curvature. Further, in many of these studies, strong mechanical inputs—such as AFM tip indentation—have been employed to drive localized deformation, limiting the ability to probe spontaneous or stable response[15–18]. Lastly, features such as crack tips have provided critical insight into nanoscale polarization confined to a few atomic layers[19,20]. Such systems present a high strain-gradient induced local curvature and, due to their strong electronic screening, ionic nature, and lack of a freestanding membrane to support flexure, these systems may obscure the intrinsic quantum mechanical origins of flexoelectricity.

Here, we explore quantum flexoelectricity in self-assembled graphene nanowrinkles (GNWr) as a structurally simple, spontaneously formed, and electronically clean system. The atomic-level buckling in these quasi-pristine structures allows direct access to extreme strain gradients, while minimizing extrinsic effects from defects, dopants, or external forcing. Together, these features enable a uniquely unambiguous exploration of flexoelectric phenomena at the nanoscale, offering a new model system to probe quantum-mechanical electromechanical coupling in low-dimensional materials.

## Results and Discussion

### A Platform for Quantum Flexoelectricity

Unlike piezoelectricity (**Fig. 1A (i)**), which arises from compressing a non-centrosymmetric crystal, flexoelectricity (**Fig. 1A (ii)**) also occurs in centrosymmetric materials under strain gradients. In GNWRs, the large quantum flexoelectric response (**Fig. 1A (iii)**) is a direct consequence of extreme curvature-driven strain gradients, pushing flexoelectricity to its upper limit within a purely carbon-based system. We achieve the formation of these GNWRs by the mechanical mismatch between two 2D materials—graphene and $MoS_2$— where the latter promotes wrinkle formation more effectively than silica due to its distinct elastic properties. (**Fig. 1B**) The incommensurate graphene-$MoS_2$ interface, resulting from their differing lattice constants (0.31 nm for $MoS_2$ and 0.24 nm for graphene), along with differences in Poisson's ratios, drives spontaneous buckling under tension. Additionally, the mismatched Young's modulus facilitates self-assembly shear-slide, further enhancing wrinkle formation. This is consistent with the 'standing-collapsed' architecture inherent to graphene under extreme strain fields (i.e., closed loops) as previously demonstrated experimentally by transmission electron microscopy and confirmed by DFT simulations[21], and also confirmed experimentally by AFM measurements in this study, see **Fig. S1**.

A key factor in this morphology is the bending modulus difference: Graphene, with a modulus (1.0 ± 0.1 TPa)[22] one order higher than $MoS_2$ (~ 270 ± 100 GPa)[23], can wrinkle under tensile strain, to maximize interaction with the substrate. The energy cost of bending is further offset by vdW interaction between the vertical walls of the GNWR. The lower interfacial shear force between graphene and $MoS_2$ compared to graphene and silica enables graphene to slide and form pronounced wrinkles, as revealed by friction force mapping (**Fig. S2**), which shows a threefold friction difference between Gr/$MoS_2$ and Gr/silica. AFM and SEM measurements indicate that GNWrs on $MoS_2$ reach heights of 6–8 nm, with a roughness ($R_q$ = 2.78 nm, $R_a$ = 1.73 nm) nearly six times higher than on silica ($R_q$ = 0.48 nm, $R_a$ = 0.28 nm) (extracted from data in **Fig. 3A**).

Unlike traditional wrinkle formation in CVD graphene due to thermal expansion mismatch, the $MoS_2$ substrate enhances wrinkle height and density in physically deposited graphene (more details in **Fig. S3**). Molecular dynamics simulations indicate this assembly is extremely rapid: small wrinkles spontaneously nucleate for local uniaxial strains $\varepsilon_x > 0.02$ and subsequently agglomerate by sliding across the $MoS_2$ within a 0.1 ns time scale. (**Fig. 1C**). Furthermore, GNWrs serve as ideal systems to evaluate the quantum flexoelectric effect, as the wrinkle tips exhibit uniform curvature longitudinally, as it is governed by a fine balance of covalent bonding, local strain, and the formation of van der Waals-stabilized lips—a structure commonly observed in graphene loops at the edges of multilayer samples. This distinction is crucial when juxtaposed with Raman data (**Fig. S4**), which shows little to no D-band signal, ruling out defect-driven effects.

Interestingly, flexoelectric strain is predominantly tensile, providing key insight into the asymmetric electronic features discussed later. For a typical high-density region of GNWrs (**Fig. 1D**), we observe up to 0.05% tensile strain, as calculated from Raman maps (**Fig. 1E**). These values, however, likely underestimate the actual strain due to the averaging effect of Raman mapping, which has a spot size of ~300 nm [24]. Graphene nanostructures exhibit a range of morphologies, including wrinkles, folds, and bubbles, all of which fall within a similar strain range. Previous work reports strain values of (0.03 ± 0.01)% for graphene bubbles and (0.04 ± 0.01)% for folded wrinkles[25,26]. More broadly, curvature-induced strain in 2D materials typically falls within the $10^{-2}$ % range[27]. Despite these comparable strain values, GNWrs occupy the upper limit of this range and exhibit a unique structural character. Unlike bubbles or folds, GNWrs possess extremely high curvature yet break symmetry in a way that induces polarization. Strain field calculations based on bond-length fluctuations reveal significant buckling under flexion near the GNWr bud of curvature $c_2$. To further investigate variations based on GNWr height, we performed DFT calculations for three different cases, now referred to as GNWr1 ($h$ = 15.64 Å), GNWr2 ($h$ = 27.54 Å), and GNWr3 ($h$ = 39.99 Å) (see **SI methods**, **Fig. S5**). **Fig. 1F** presents strain gradient values for a DFT-relaxed GNWr, mapping bond-length variations in the vicinity and away from the GNWr (GNWr1 and GNWr2 in **Fig. S6**).

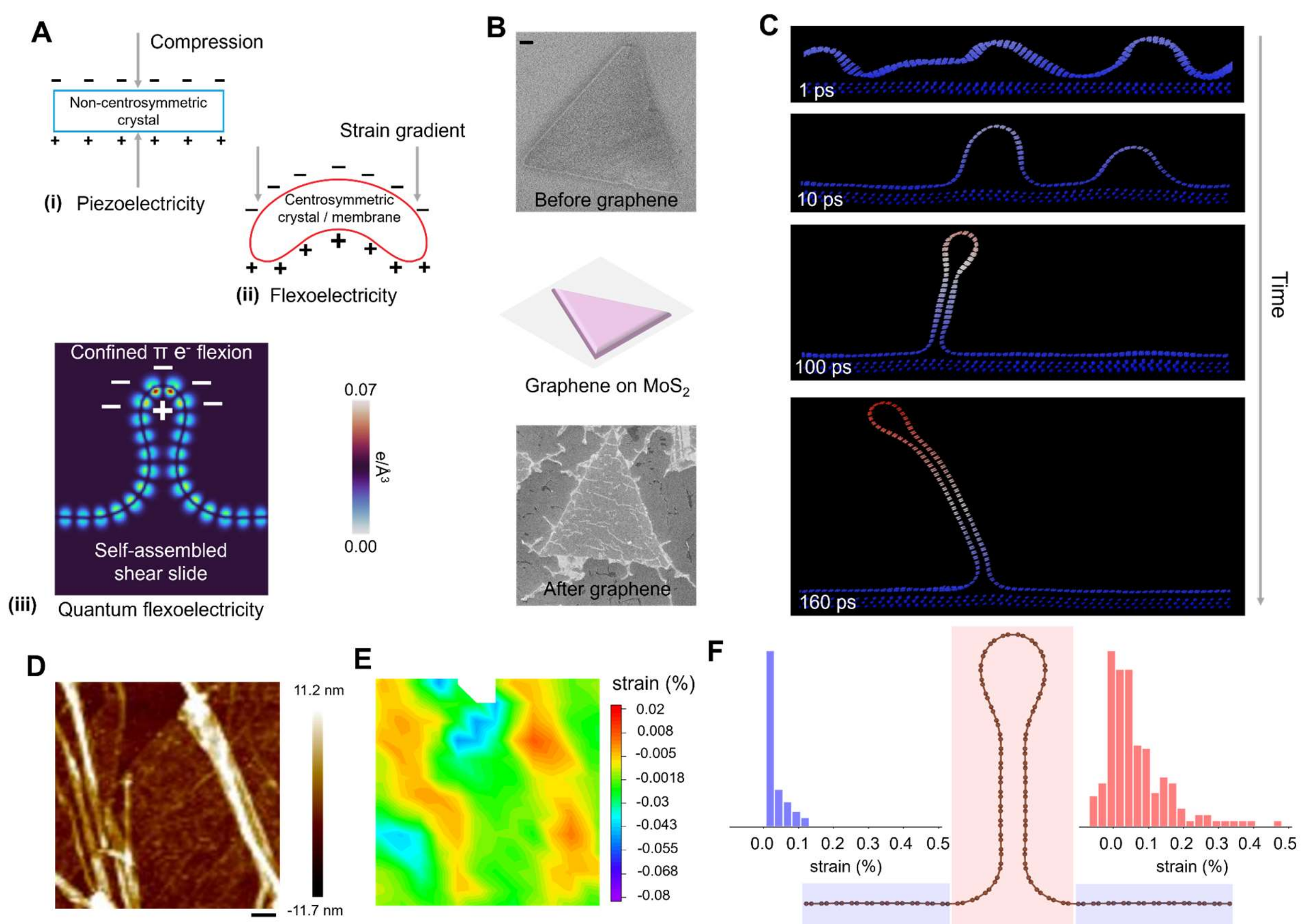

**Figure 1. A platform for quantum flexoelectricity.** (A) Schematic illustration comparing (i) piezoelectricity, (ii) flexoelectricity, and (iii) quantum flexoelectricity based on crystal centrosymmetry, methods of compression/strain, mechanisms of relaxation, and curvature. The charge density in (iii) is obtained by DFT and shows the density in a 2 eV window below the Fermi energy. (B) Scanning electron microscopy (SEM) micrographs of a flat $MoS_2$ substrate before and after the transfer of monolayer graphene and subsequent GNWr self-assembly. The scale bar is 1 µm. (C) Molecular dynamics simulations depicting the time evolution of GNWr self-assembly at T = 100 K. Smaller wrinkles spontaneously nucleate under compressive strains $\varepsilon_x > 0.02$, and rapidly combine into large, standing collapsed wrinkles within a 0.16 ns time scale. (D) AFM topology and corresponding (E) Strain map obtained from Raman spectroscopy maps (F) Histogram of calculated lattice strain across a GNWr. The scale bar is 200 nm.

**Sub-microscopic manifestations of sub-nm flexoelectricity.**

The resulting GNWrs form a curvature-driven ripple, exhibiting a net polarization leading to the relationship:

$$P = f(c_1 + c_2) \tag{1}$$

where P is the induced polarization, $f$ is the material-dependent flexoelectric constant, and $c_1$ and $c_2$ are the principal Gaussian curvatures on the surface. Here, the first Gaussian curvature in the plane of the graphene sheet (along the ripple), $c_1$, is ~ 0, and the second Gaussian curvature in the perpendicular direction, $c_2$ = 1/R, where R is the radius of curvature, when we approximate the tip of the GNWr as roughly circular, as depicted in **Fig. 2A**) therefore:

$$P = f/R, \tag{2}$$

showing that the polarization effect is strongest in materials with a large flexoelectric constant and a small local radius of curvature. In bulk perovskites, for example, $f$ is typically on the order of ~ nC/m[28] , whereas in 2D systems, values can vary widely depending on material properties and strain conditions. For graphene, DFT predicts f ~ 2 nC/m[6]. While direct measurements of $f$ in graphene are challenging, our experimental results and observed polarization response closely match DFT predictions, confirming the estimated value of $f$.

What distinguishes nanomaterials is not necessarily a large flexoelectric constant—no greater than that found in biological membranes[29]—but rather the ability to achieve nanometer-scale radii of curvature. In carbon nanostructures, a reasonable lower limit for curvature is the radius R~3.4 Å as seen in $C_{60}$ fullerenes or (5,5) and (9,0) nanotubes. However, despite their high curvature, these closed structures do not exhibit net polarization due to their symmetric assembly—each local curvature-induced dipole, while potentially large[30] is canceled by an opposing antiparallel dipole, leading to a net zero polarization. This curvature-induced charge separation across the walls of carbon nanotubes was predicted in 2002 using DFT[31]. In contrast, the GNWrs considered here exhibit a local curvature corresponding to an estimated radius of ~5 Å[21], approaching the lower limit, yet without a canceling effect. This unique structural asymmetry allows for a net flexoelectric response, distinguishing GNWrs from other carbon-based nanostructures.

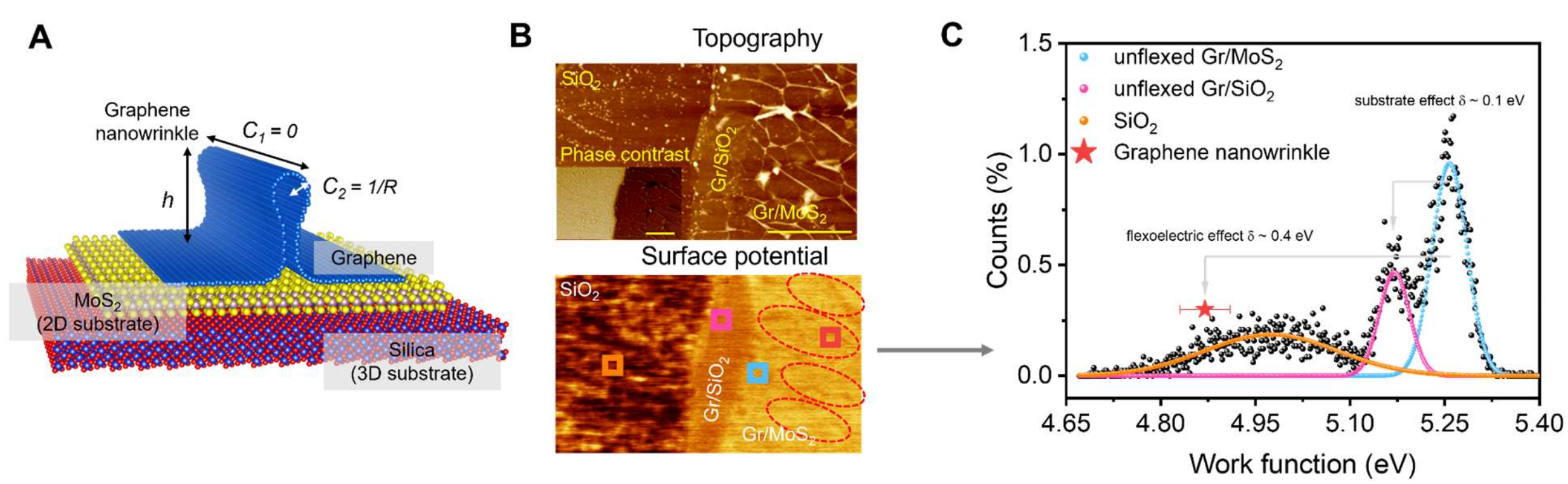

**Figure 2. Flexoelectric effect-driven work function shifts.** (A) Schematic representation showing the carbon GNWr (blue) on $MoS_2$ (yellow/grey), on top of $SiO_2$ substrate (red and dark blue). The schematics show the Gaussian curvatures in the GNWr, utilized to establish the giant flexoelectric formalism. (E) Atomic force microscopy (AFM) is used for topography and phase contrast, with corresponding Kelvin probe force microscopy (KPFM) measuring surface potential across different substrate regimes. (F) Work function distribution across surfaces, highlighting the significant shifts observed in GNWrs compared to expected values. All scale bars are 1 um.

It follows that the emerging polarization can lead to measurable effects on the electronic properties. For instance, the pronounced work function shift observed in our study underscores the dominant role of the flexoelectric effect over any substrate-driven influence. This is evident in the significant work function shift as measured by Kelvin Probe Force Microscopy (KPFM) **(Fig. 2B-C)**, where GNWrs (~4.9 eV) exhibit values closer to pure silica (~4.97 ± 0.28 eV) than unflexed graphene on $MoS_2$ (~5.27 ± 0.1 eV) or unflexed graphene on silica (~5.17 ± 0.1 eV). If $MoS_2$ were to influence the electronic structure through substrate-driven work function shifts, this effect would be uniform across the entire material rather than localized to GNWr. Instead, the observed work function shift is highly confined to flexed regions (where the radius of curvature is smallest, see **eq 2**), demonstrating that the flexoelectric effect (δ ~ 0.4 eV) is far stronger than any global substrate-mediated doping in unflexed graphene (δ ~ 0.1-0.2 eV) (**Fig. 2B**). The stark deviation of GNWr work function values from those of unflexed graphene further confirms that substrate interactions play a negligible role in modulating local charge distribution. Instead, the observed shift originates from flexoelectric-induced polarization.

From experiment and theory, we establish that the flexoelectric effect is also independent of the GNWr height, making it a reliable method for creating high-density wrinkle networks that behave identically. This also highlights the central role of the localized curvature at the top of the wrinkle, further confirming that the observed phenomena are indeed due to flexoelectricity. To probe direct manifestations of the effect, in the form of in-plane flexoelectric current, we performed conductive atomic force microscopy (c-AFM). If we consider a dense network of GNWrs **(Fig. 3A)**, with features of varying heights, when a fixed bias (2 V) is applied (**Fig. 3B)**, all GNWrs yield a similar value of flexoelectric current (~67 pA) (**Fig. 3C**).

The results show that despite no bandgap opening (**Fig. S7**) the flexoelectric effect is preserved regardless of height (**Fig. S8**), confirming that the curvature at the GNWr tip is the driving force behind the observed phenomena. This is further consistent with **Fig. S9**, revealing that the field induced by charge reorganization in response to an external field (depolarization) highlights the key role of the GNWr apex. Comparing features, such as GNWrs and bubbles, the role of curvature is clearly dominant— not the height of the feature— as measurements independently show in **Fig S10**.

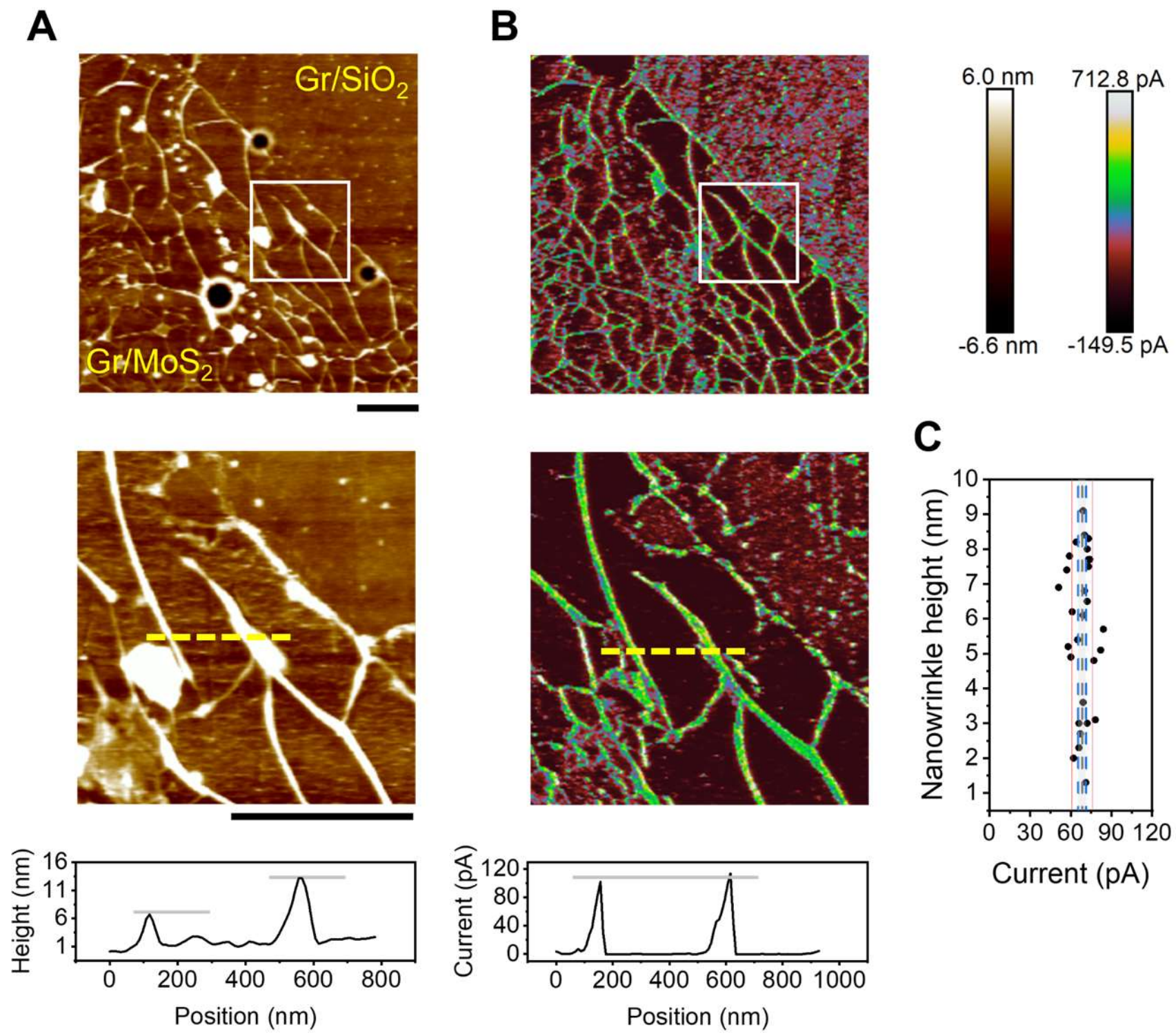

**Figure 3. Large-area, uniform, quantum flexoelectricity.** (A) Atomic force microscopy (AFM) and (B) corresponding conductive AFM characterization of the topology and current distribution of GNWr. Interestingly, only the GNWrs exhibit a higher baseline current, which is independent of GNWr height (i.e., independent of height *h*), as shown in (C). All scale bars are 1 μm.

**Tuning quantum flexoelectricity near fundamental curvature limits.**

To further investigate the flexoelectric response, we performed c-AFM measurements, now, under varying bias conditions and looping the spatial coordinates (**Fig. 4A**). As observed previously, and confirmed by our DFT calculations, the effect depends on the applied bias and remains uniform for a given bias, regardless of GNWr height, resulting in large-area coverage and reproducibility, as seen in **Fig. 4B**. Dipole moments calculated for these systems were found to be 0.0261 e Å (0.125 D), 0.0223 e Å (0.107 D), and 0.0297 e Å (0.143 D), respectively (see **Fig. S11** for fitting details). This shows minimal variation across various sizes of GNWrs. **Fig. 4C** analyzes four specific GNWrs, looping the same spatial resolution across different voltages, and reveals a clear asymmetry in response when probing low and negative biases. In this bias regime, a threshold voltage ($\Phi_{th}$) of ~1.01 V is observed (**Fig. 4D**), corresponding to a barrier potential that must be overcome, consistent with the flexoelectric dipole induced in the GNWr. The induced polarization results in a significant local work function shift in the vicinity of the wrinkles compared to flat graphene, as predicted by DFT (**Fig. 4E**). This work function difference leads to a band offset at the interface between wrinkled and flat graphene, forming a type-I heterojunction, with the calculated offset (~1.2 V) shown in **Fig. 2F**. Together, with the fact that there is no bandgap opening (**Fig. S7**) these results demonstrate the influence of strain gradients on electron distribution and electrostatic properties within confined 2D volumes.

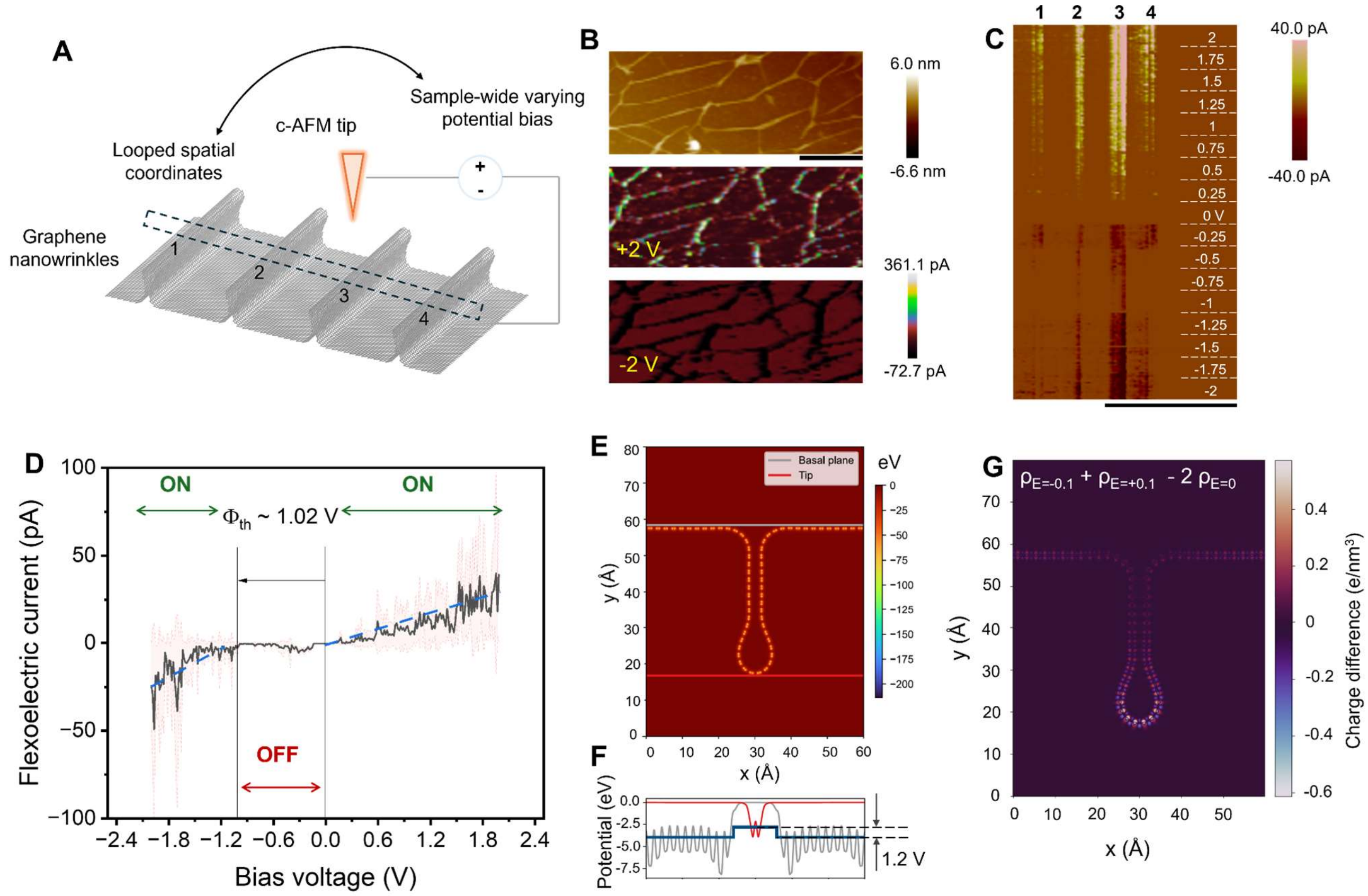

**Figure 4. Potential bias dynamics of flexoelectrics.** (A) Conductive Atomic Force Microscopy (CAFM) measurement setup to study flexoelectric response in GNWr (B) spatially-resolved CAFM data under fixed opposite biases show asymmetry in current response (C) Bias voltage-dependent data where each 'strip' consists of a repeated area over the sample under a different potential. Numbers indicate the position of 4 GNWr. At zero bias, the GNWr's flexoelectric signatures are not observed and come to being when bias is applied. (D) Flexoelectric current measurements were collected specifically on GNWr and averaged across a bias range. (E) Electrostatic potential map of a GNWr with potential extracted at the tip and basal plane, as shown in (F). (F) also shows the macroscopic average potential (blue), showing that the flexoelectric polarization yields a reduction of the work function at the wrinkle tip and the ensuing band alignment. (G) Asymmetry in electrostatic response obtained by computing charge density difference between opposite and equal magnitude potential fields. The presence of flexoelectric polarization is responsible for the asymmetry in the charge distribution upon application of opposite electric fields. All scale bars are 1 µm.

Such an asymmetric response under bias conditions is due to the net flexoelectric-induced polarization (**Fig. 4G**); and highlights the potential for using these structures as densely packed, high-density switches, as shown in **Fig. 4D**, for advanced nanoscale electronic applications. Our observed effect represents the most pronounced flexoelectric response to date, driven by exceptionally large curvature—approaching the fundamental limits of bond lengths—and stands as the only experimental demonstration in low-dimensional material systems thus far, as shown in **Fig. 5A**. As discussed previously in Eq. 2, there is a strong inverse relationship between the radius of curvature and the flexoelectric effect. Using $\Phi_{th} \sim 1$ V from **Fig. 4D**, we can describe it as the potential of a dipole as measured at proximity as follows:

$$\Phi_{th} = \frac{1}{4\pi\varepsilon_0}\frac{p}{r^2}\cos\theta \quad (3)$$

where $p$ is the dipole moment, $r$ is the separation between the GNWr and scanning probe microscopy tip (~ 2.52 ± 0.27 nm, as determined in a separate experiment; see **Fig. S12**), and $\theta$~ 0 due to normal incidence during measurement. Extracting the polarization density from eq (2) and (3) for theory and experiment yields *giant* magnitudes $P_{th}$ ~ 4 C/m$^2$, $P_{exp}$ ~ 1 C/m$^2$. Given our curvature of 5 x 10$^{-10}$ m$^{-1}$, we move towards fundamental limits for these values, as shown in **Fig. 5B.**

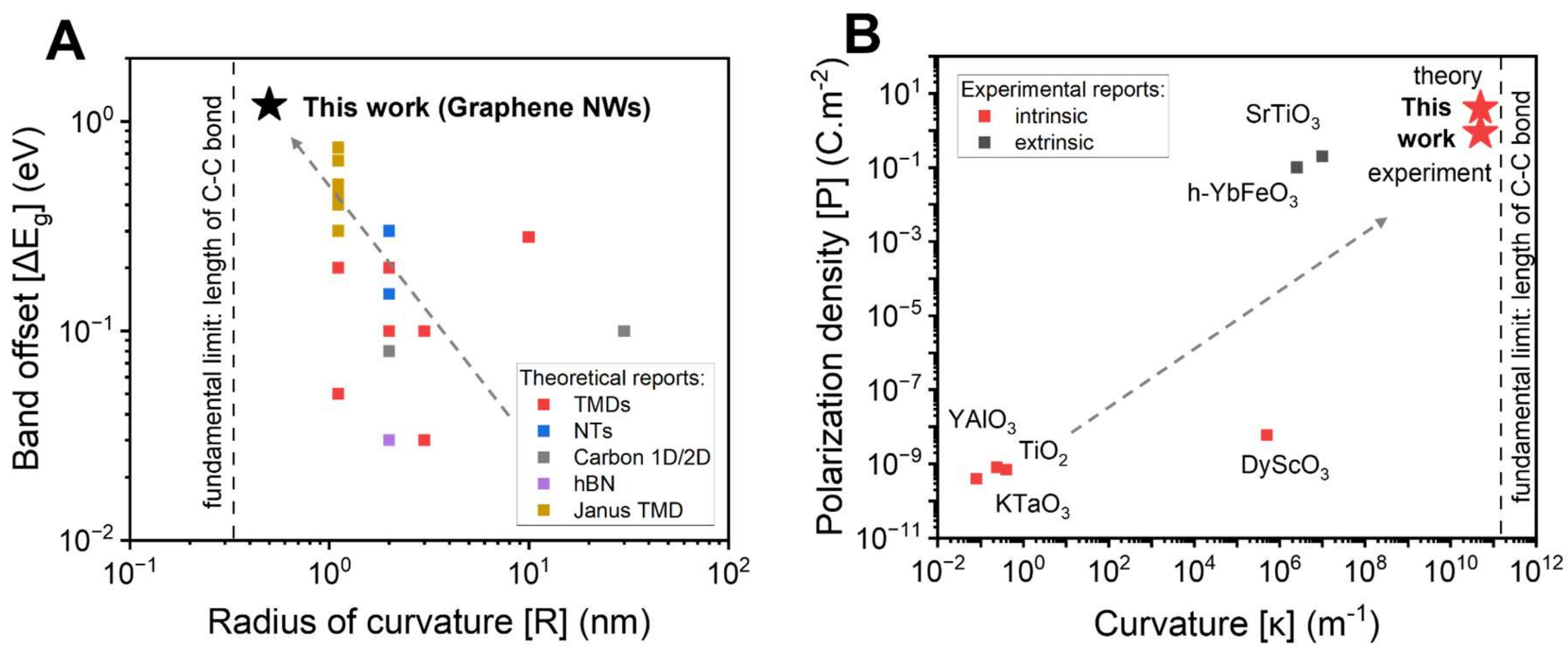

**Figure 5. Quantum flexoelectricity approaches fundamental limits of polarization densities.** (A) Comparing band offsets in different 2D material systems as a function of flexoelectric radius of curvature. Note: all reports curated, except for this work, are only theoretical reports[30,32–36]. (B) Comparing polarization density as a function of flexoelectric curvature. Note: all reports curated, except for this work, are only experimental reports on bulk materials[16,28,37,38]. Trend lines are linear on a log-log plot, indicating power law features scaling for curvature.

Our analysis further reveals that the curvature at wrinkle termini can exceed that of engineered structures by several orders of magnitude, significantly amplifying strain gradients and flexoelectric effects. The transition from flat regions to high-curvature domains provides optimal conditions for investigating the relationship between geometry and flexoelectric response, expanding insights for future device design[39]. The combination of high reproducibility, scalability, and extreme curvature makes graphene nanowrinkles particularly promising for applications in energy harvesting, flexible electronics, and strain-tunable quantum devices. The ability to induce strong flexoelectric polarization without external gating suggests potential use in non-volatile memory and strain-gated transistors. Furthermore, the self-assembled nature of these wrinkles provides a cost-effective pathway for scalable flexoelectric device fabrication, which could be integrated with lab-to-fab approaches for scaling up production[40]. Future work should explore integrating GNWrs into functional electronic architectures, as well as investigating potential coupling effects with other 2D materials to engineer hybrid[41,42] flexoelectric systems.

**Acknowledgments:** We thank the Rice Shared Equipment Authority (SEA) for their support.

**Funding:** SAI and PMA acknowledge and thank the Quad Fellowship. MT and ABD acknowledge the Sussex Strategy Development Fund. JGM is supported by the University of Manchester Dame Kathleen Ollerenshaw Fellowship.

**Author contributions:** SAI, VM, and MT conceived the project and designed the research. SAI, VM, and MT collected, analyzed, and presented the data and co-wrote the manuscript through the contributions of all authors. JGM performed the MD simulations. VM performed the DFT calculations. JPS collected scanning electron micrographs. MT, AD, RV, VM, and PMA acquired funding for the project. VM, PMA, AD, and MT supervised the entire project.

**Competing interests:** The authors declare that they have no competing interests.

**Data and materials availability:** Data are available in the manuscript or supplementary materials and raw data are available upon request to the authors.

**Supplementary Materials**

Summary

Methods

Supplementary Videos S1-7

Figs. S1 to S12

**Supplementary Information**

# Sub-nm Curvature Unlocks Quantum Flexoelectricity in Graphene

Sathvik Ajay Iyengar[1], James G. McHugh[2], Jonathan P. Salvage[3], Robert Vajtai[1], Alan Dalton[4]*, Manoj Tripathi[4, 5]*, Pulickel M. Ajayan[1]*, Vincent Meunier[6]*

[1] *Department of Materials Science and NanoEngineering, Rice University; Houston, TX, 77005 USA*

[2] *National Graphene Institute, University of Manchester, Booth St. E., Manchester, M13 9PL, United Kingdom*

[3] *School of Pharmacy and Biomolecular Science, University of Brighton, Brighton BN2 4GJ, U.K*

[4] *Department of Physics and Astronomy, School of Mathematical and Physical Sciences, University of Sussex; Brighton BN1 9QH, United Kingdom*

[5] *Department of Civil and Environmental Engineering, South Dakota School of Mines and Technology, Rapid City, SD 57701*

[6] *Department of Engineering Science and Mechanics, Pennsylvania State University; University Park, PA, 16802, USA*

*Corresponding author. Email: Alan Dalton A.B.Dalton@sussex.ac.uk, M.Tripathi@sussex.ac.uk, ajayan@rice.edu, vincent.meunier@psu.edu

**This file includes:**

- Methods
- Figs. S1 to S12
- References
- Supplementary movies 1-7 (attached)

## Methods:

### Sample fabrication:

The $MoS_2$ layer was grown via chemical vapor deposition (CVD) on a silica substrate. Comprehensive synthesis methodology can be found elsewhere[43,44]. CVD-grown graphene on Cu substrate was purchased from *Graphenea* (Spain), with added assistance related to wet transfer onto the $MoS_2$/silica surface. The heterolayer structure was treated in an acetone bath at 40°C for 3 hours to remove PMMA residue.

### Raman spectroscopy:

Raman spectroscopy is carried out by Renishaw inVia™ confocal Raman microscope with 0.8 $cm^{-1}$ spectral resolution. 532 nm laser (type: solid state, model: RL53250) at a power of 5 mW with 1800 $mm^{-1}$ grating in 100x magnification is used. The peak position and peak intensity are then fitted by Lorentz fitting to monitor the Raman peak shift ($cm^{-1}$).

**Raman-strain calculations:**

Using these peak shift Strain (%) has been measured using relation (s1):

$$\begin{pmatrix} \omega_1 \\ \omega_2 \end{pmatrix} = T \begin{pmatrix} \varepsilon \\ \eta \end{pmatrix} \quad \text{… (s1)}$$

where,

$$T = \begin{pmatrix} -2\gamma_1 \omega_1^0 & k_1 \\ -2\gamma_2 \omega_2^0 & k_2 \end{pmatrix} \quad \text{… (s2)}$$

*(γ)* is the Grüneisen parameter, *(K)* is the doping shift rate, and *(ω°)* is the no-strain and no-doping peak position taken from the reference sample where graphene is suspended over a lithography-etched textured surface. The subscript denotes the corresponding Raman modes. As for graphene, $\omega_1$ and $\omega_2$ are G and 2D modes, where ($\gamma_G$) = 1.95, ($\gamma_{2D}$) = 3.15, $k_G = -1.407 \times 10^{-12}$ $cm^{-1}$, and $k_{2D} = -0.285 \times 10^{-12}$ $cm^{-1}$(*27*, *41*).[26,45] The vector space of Raman peak positions $\omega_1$-$\omega_2$ is a linear transformation from the ε-n space, while the origin of both spaces defines the absence of strain and doping. Therefore, ω represents the deviation of the recorded frequency from ω° due to strain or doping. Notably, *(n)* represents the relative shift in the charge carrier and mostly originates from the charge exchange with the substrate. Also, the airborne impurities adsorb over the surface and at the edge region may influence *(n)*.

**Scanning probe microscopy:**

Atomic Force Microscopy (AFM) characterization was carried out using a Bruker Dimension Icon with PF-QNM (PeakForce-Quantitative NanoMechanical) mode and friction mode, simultaneously measuring topography, adhesion, and friction force. Conductive AFM (C-AFM) was performed using PF-TUNA mode with a conductive probe (PFTUNA, a silicon nitride cantilever coated with Pt/Ir). The cantilever has a stiffness of 0.45 N/m. The conduction setup was established by connecting the top graphene layer to the AFM stage using conductive tape. C-AFM measurements were conducted at bias voltages ranging from -2 V to 2 V.

Kelvin Probe Force Microscopy (KPFM) was performed at room temperature using a Bruker AFM (model: ScanAsyst) with a PFQNE-AL cantilever. A single layer of CVD $MoS_2$ was used for the investigation, treated with different analytes, and its surface potential was measured. The KPFM measurement followed the two-pass technique, where the first pass scanned the sample's topography, and the second pass, conducted at a specific lift height, measured the contact potential difference (CPD, mV).

A reference sample (Au–Si–Al) was used to calibrate the cantilever probe apex and determine the sample's work function (eV).

**Theory:**

**Molecular Dynamics:**

Molecular dynamics (MD) simulations were carried out in the LAMMPS MD environment[46]. In-plane bonding of $MoS_2$ was approximated using a reactive empirical bond order potential parameterization[47], whereas graphene was simulated with the AIREBO-M potential[48]. The latter adapts long-range C–C interactions to a Morse potential, fitted to quantum chemistry data, making it particularly suitable for high pressures and strains. Interlayer interactions between graphene and $MoS_2$ were approximated by Lennard-Jones 12/6 potentials for C-Mo and C-S interactions; parameters were initially taken from the Lorentz-Berthelot mixing rule[49], and were re-optimized to fit DFT-calculated interlayer adhesion. An additional, Lennard-Jones 9/3 ($\sigma$ =4 Å, $\varepsilon$ = 6 meV) confining wall is applied underneath the $MoS_2$ layer, to mimic the effect of an underlying substrate. Periodic boundary conditions are applied in the in-plane directions.

| Pair | σ (Å) | ε (meV) |
|---|---|---|
| C-S | 3.3 | 4.657 |
| C-Mo | 4.5 | 1.826 |

**Table S1:** DFT-parameterized Lennard-Jones parameters for C–S and C–Mo interactions.

<u>Wrinkle nucleation:</u> Finite-temperature simulations of graphene/$MoS_2$ heterostructures were performed in a periodic "ribbon" geometry, where monolayer segments of graphene and $MoS_2$ are lattice-matched along the shorter ($y \approx 7.5$ nm) direction to eliminate strain. Increasing quantities of additional graphene were then added in the long ($x \approx 32.5$nm) direction, imposing an initial, uniform uniaxial strain ($\varepsilon_x$) in the graphene layer, mimicking slip and compressive strain of excess material over the $MoS_2$ surface. Hetero-bilayers were then simulated using an NVT ensemble at T = 100 K over short, 0.06 ns timescales. Importantly, for small amounts of additional material (initial uniaxial strains of $\varepsilon_x \lessapprox 0.02$), the graphene layer remains uniformly compressed without developing wrinkles. This behavior is consistent with the fact that spontaneous wrinkling, driven by the softening of flexural modes, only emerges once the compression exceeds a critical threshold[50]. With increasing values of initial uniaxial strain, multiple, small wrinkles nucleate, and can rapidly agglomerate within the simulation time scale due to the incommensurate lattice structure and low adhesive energy of the graphene/$MoS_2$ heterointerface (see **Supplementary Movies 1-4**), which allows neighboring wrinkles to combine through mass transfer, by gliding across the $MoS_2$ surface (e.g. **Supplementary Movie 5**). Above wrinkle heights of ~2 nm, adhesion between nanowrinkle walls causes it to adopt a standing collapsed structure within the simulation time scale (**Supplementary Movie 6**).

Wrinkle structure and height: Additional simulations were also performed in a larger simulation cell ($x \approx 130$ nm), to assess the dependence of wrinkle geometry on height. A representative example is shown in **Supplementary Movie 7**, which shows the rapid growth of a collapsed wrinkled within a 0.16 ns timescale. The collapsed wrinkle reaches a free-standing height of $h_c \approx 8$ nm, at which point it starts to bend and tilt towards the flat graphene surface, suggesting a transition to folded structures above $h_c$.

***Ab initio* DFT:**

Density Functional Theory (DFT) was used to calculate the relaxation of charges and the deformation of charge clouds under the influence of an electric field. We used GPAW[51,52] with PBE[53] for the exchange-correlation potential with a Grimme correction for vdW interactions[54]; 1x1x10 k-point Monkhorst-Pack grid[55], and an LCAO basis (DZP). We considered three different structures with varying ripple heights (see **Fig. S5**) The structures were relaxed until forces were all below 0.05 eV/Å. We note that PBC was not applied in the direction of the applied electric field.

Determination of flexoelectric constant: For this calculation, there is a challenge that dipole moments are not well defined in periodic systems. For this reason, we used a linear response approach where the energy E upon bending and in an external electric z field is given by:

$$E = E_0 + \mu \, \varepsilon_z + P \, \varepsilon_z^2 + O(\varepsilon_z^3) \qquad \ldots \text{(s3)}$$

where $E_0$ is the energy without an electric field, μ is the dipole moment, and P is related to the polarizability. $\varepsilon_z$ is the static response due to bending, with P being the dynamic response to the electric field. Note that in the equation above, we limited the study to the linear case (that is: we only consider the field in one direction, perpendicular to the graphene layer) otherwise $\mu$ would be a vector and P would be a tensor. For this calculation, we performed self-consistent DFT calculations for 33 values of the electric field from -0.15 to 0.15 V/Å. We then performed a fit for each structure and obtained the dipole and polarizations. An example of such output is shown in **Fig. S11**.

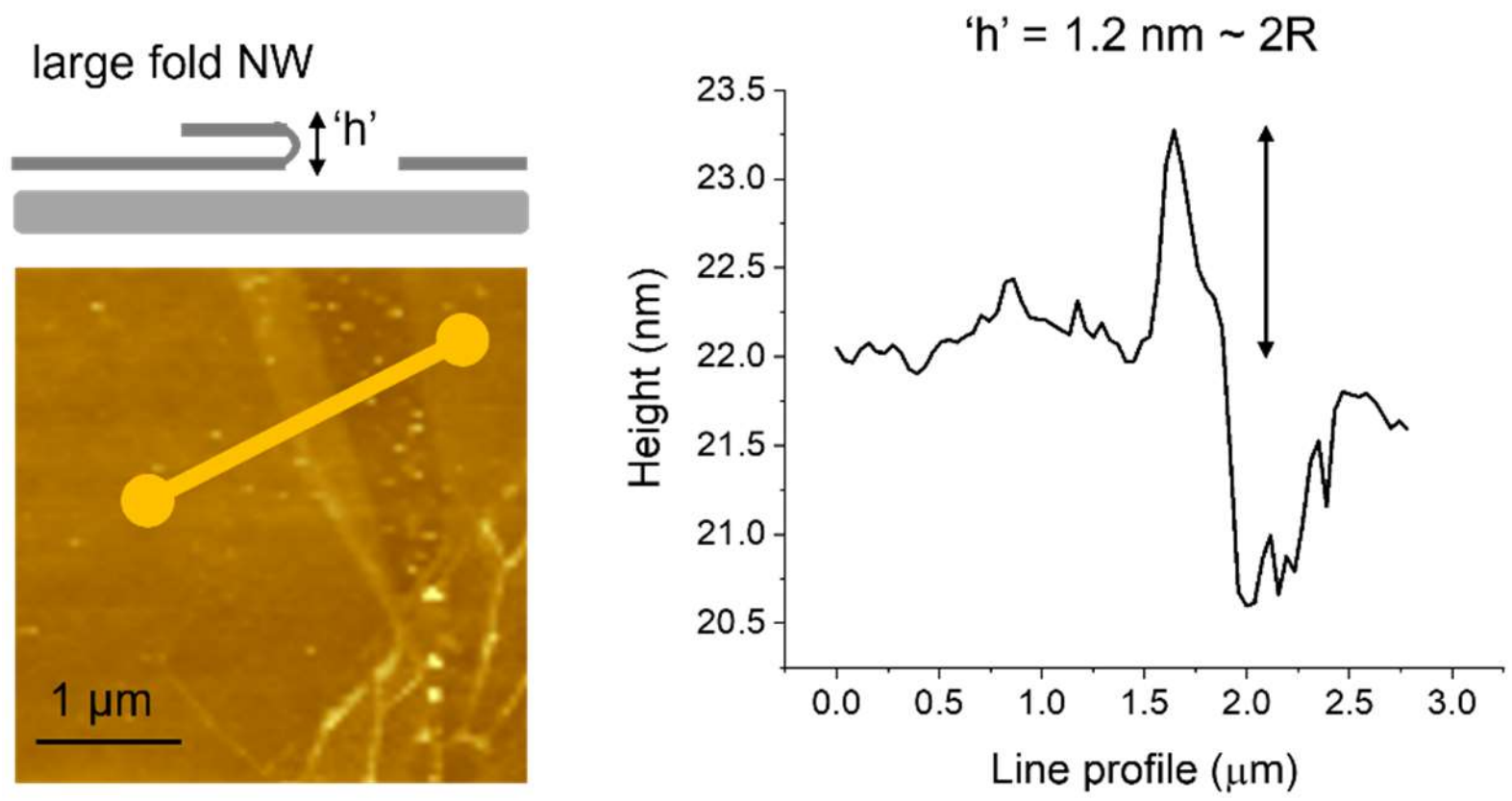

**Figure S1:** Atomic force micrograph (bottom-left) of a large fold (schematic in top-left) and corresponding height profile (right) indicates a height 'h' of roughly 1.2 nm.

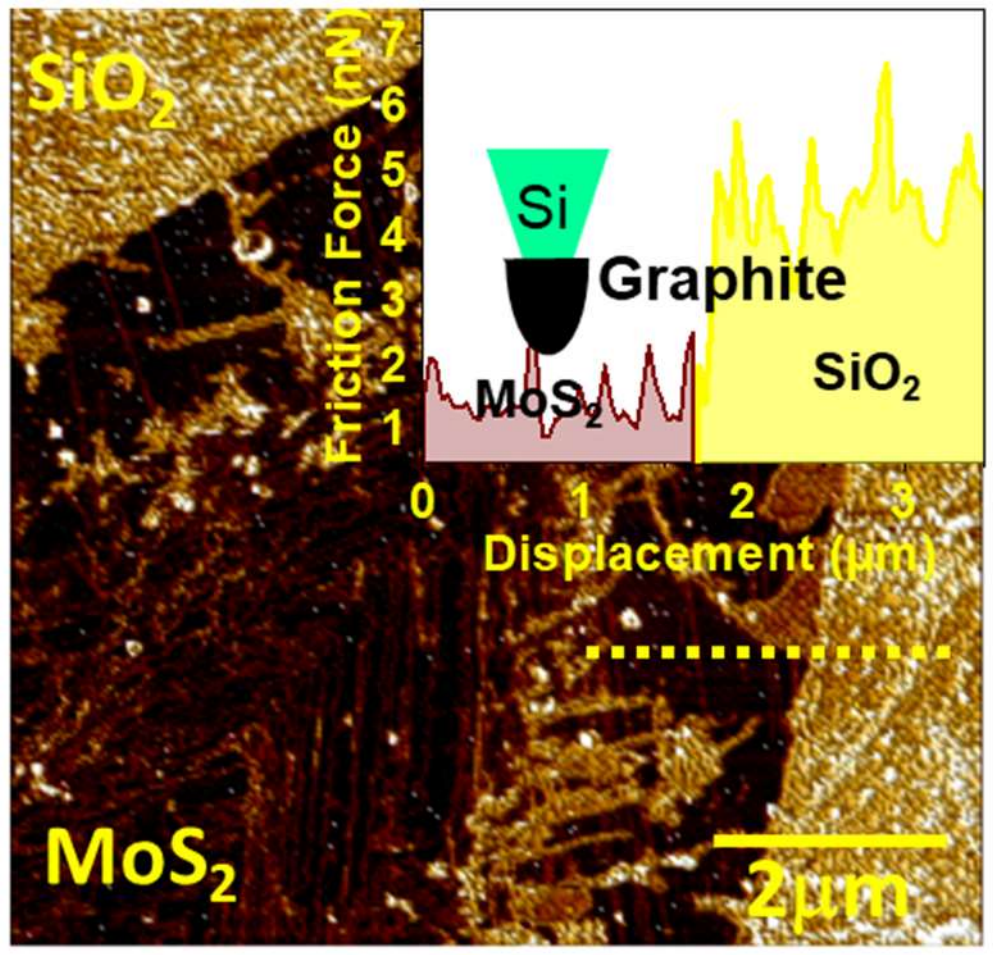

**Figure S2:** Friction force microscopy using a graphene-functionalized probe over $MoS_2$ and silica surfaces. The color contrast represents distinguishable friction force values, with the $MoS_2$ interface exhibiting a threefold lower friction force than silica under the same operating conditions (i.e., applied load and probe velocity).

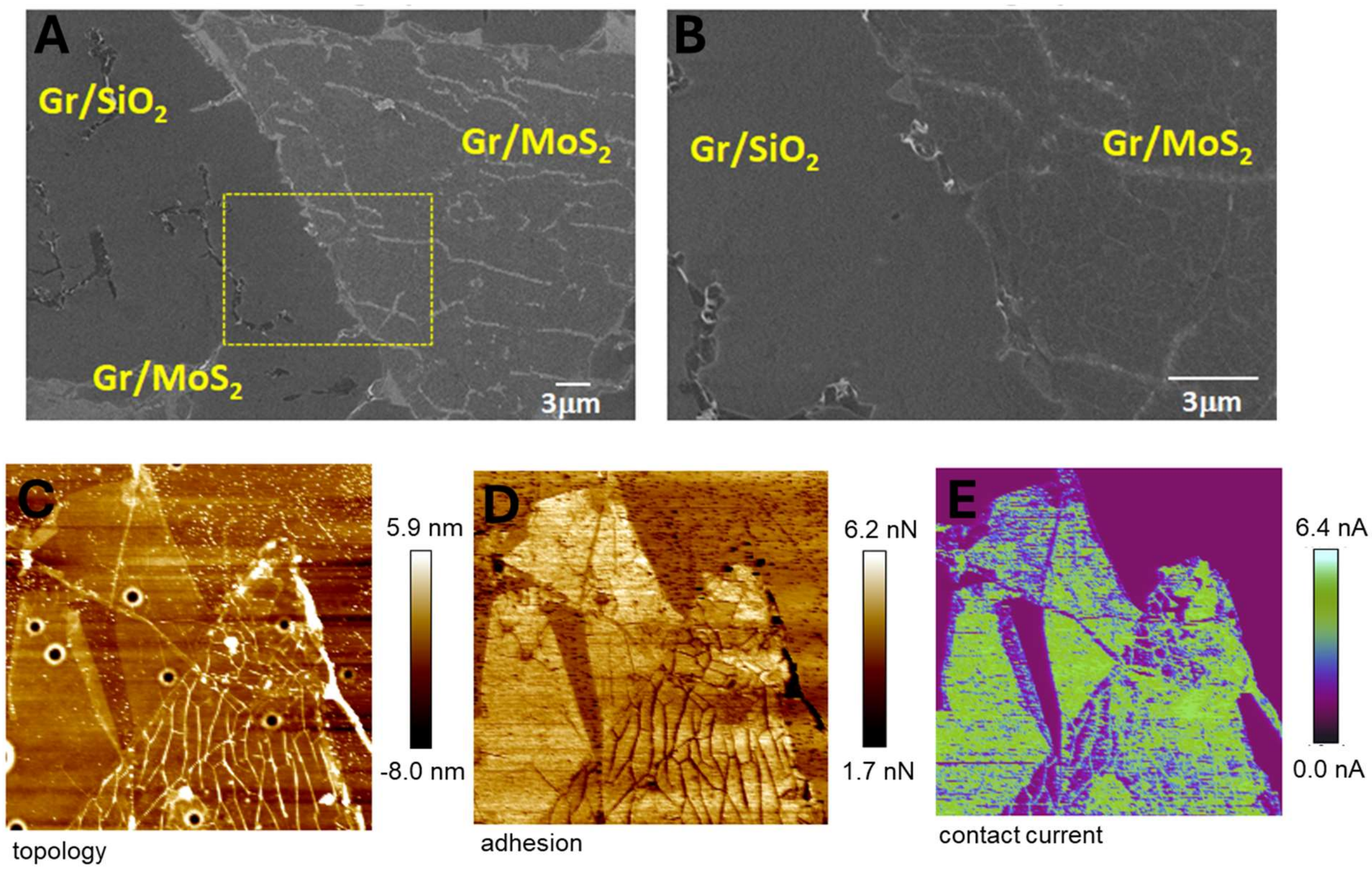

**Figure S3:** (a) SEM micrographs of graphene-covered silica and $MoS_2$ reveal a high density of graphene wrinkles on the $MoS_2$ surface. The brighter SEM contrast in wrinkled and corrugated graphene arises from variations in electron scattering trajectories, leading to higher intensity within the same area. (b) High-resolution SEM micrograph showing a relatively flat graphene surface on silica compared to the more corrugated structure on $MoS_2$. (c) AFM topography of graphene-covered silica and $MoS_2$ supports the SEM results, confirming the higher density of graphene wrinkles on $MoS_2$. (d) The corresponding adhesion force (pull-out) map reveals surface chemistry differences between the graphene-covered region and the bare silica substrate, showing that flat graphene exhibits higher adhesion force (nN) values toward AFM probes compared to wrinkles. (e) The conductive map of the same region, obtained under similar operating conditions, shows the current (nA) distribution across the graphene layer on silica and $MoS_2$ substrates. Notably, graphene wrinkles on $MoS_2$ exhibit higher conductivity than extended graphene layers on silica.

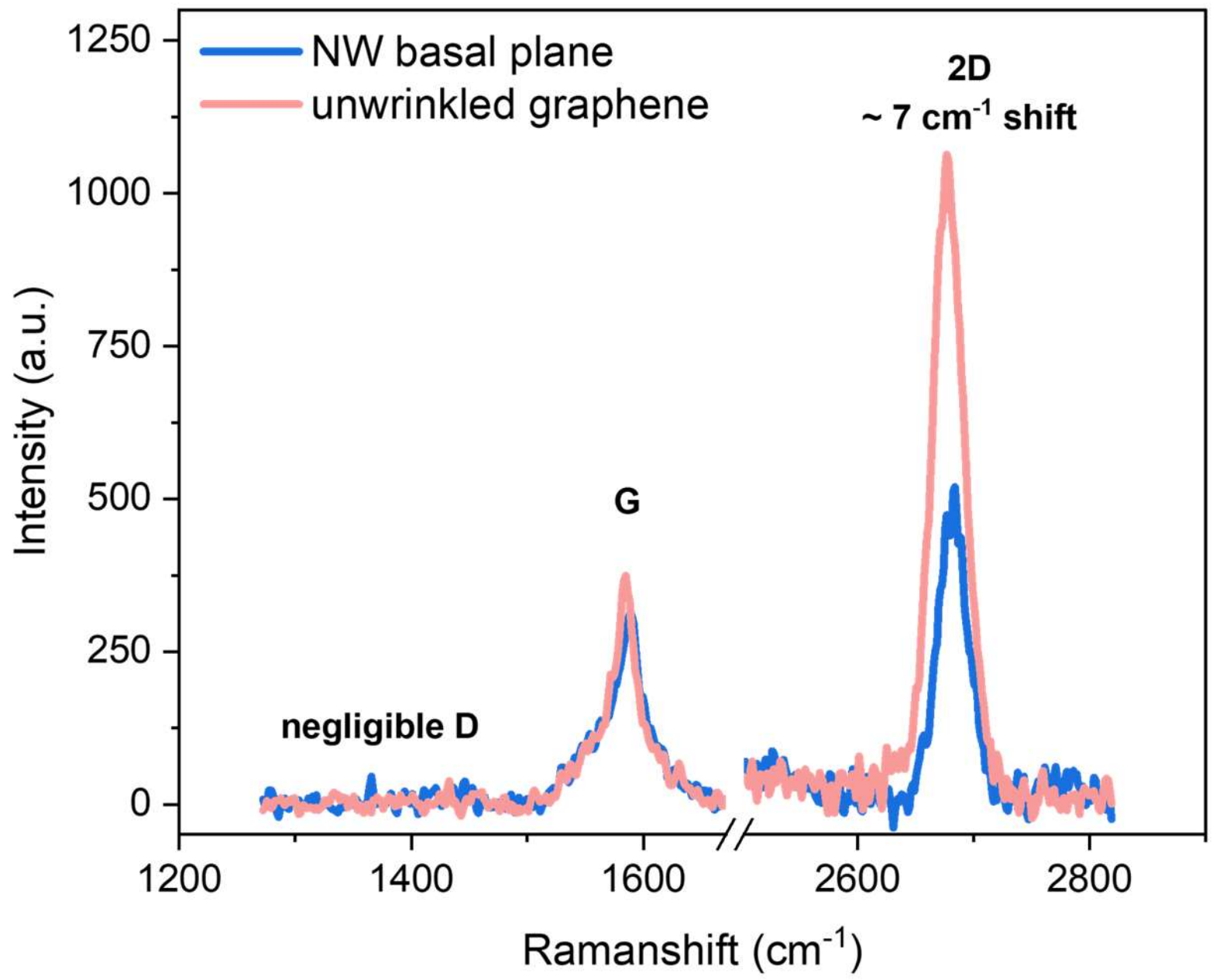

**Figure S4:** Raman spectra of graphene (D, G, and 2D) at the GNWr basal plane and when it is unwrinkled. There is a red shift in G peak position ($\omega_1$ = 4.9 cm$^{-1}$) and 2D peak position ($\omega_1$ =7 cm$^{-1}$) at GNWr as compared to basal plane.

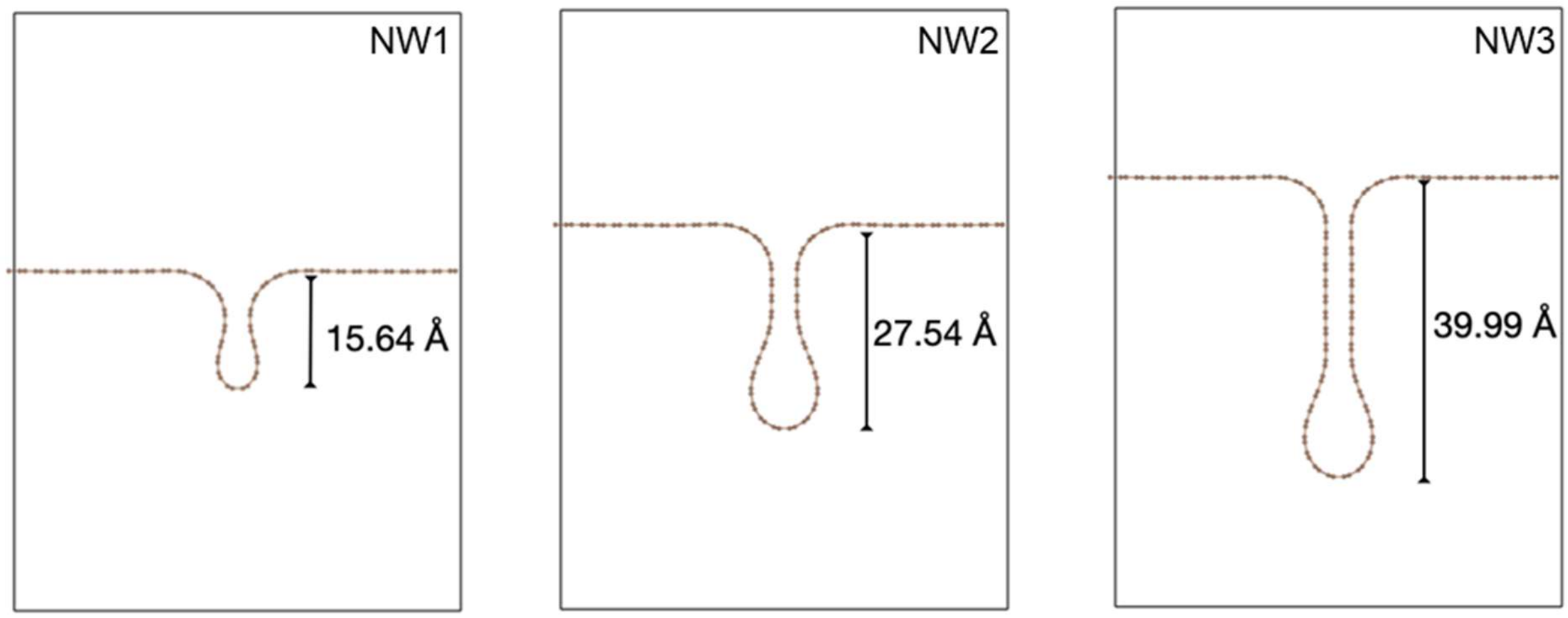

**Figure S5:** DFT relaxed structures of three wrinkle systems considered. In our main studies, the ~4 nm (GNWr3) wrinkle model is used. For relaxation, a flat and rigid layer was added to the top to mimic deposition on a substrate.

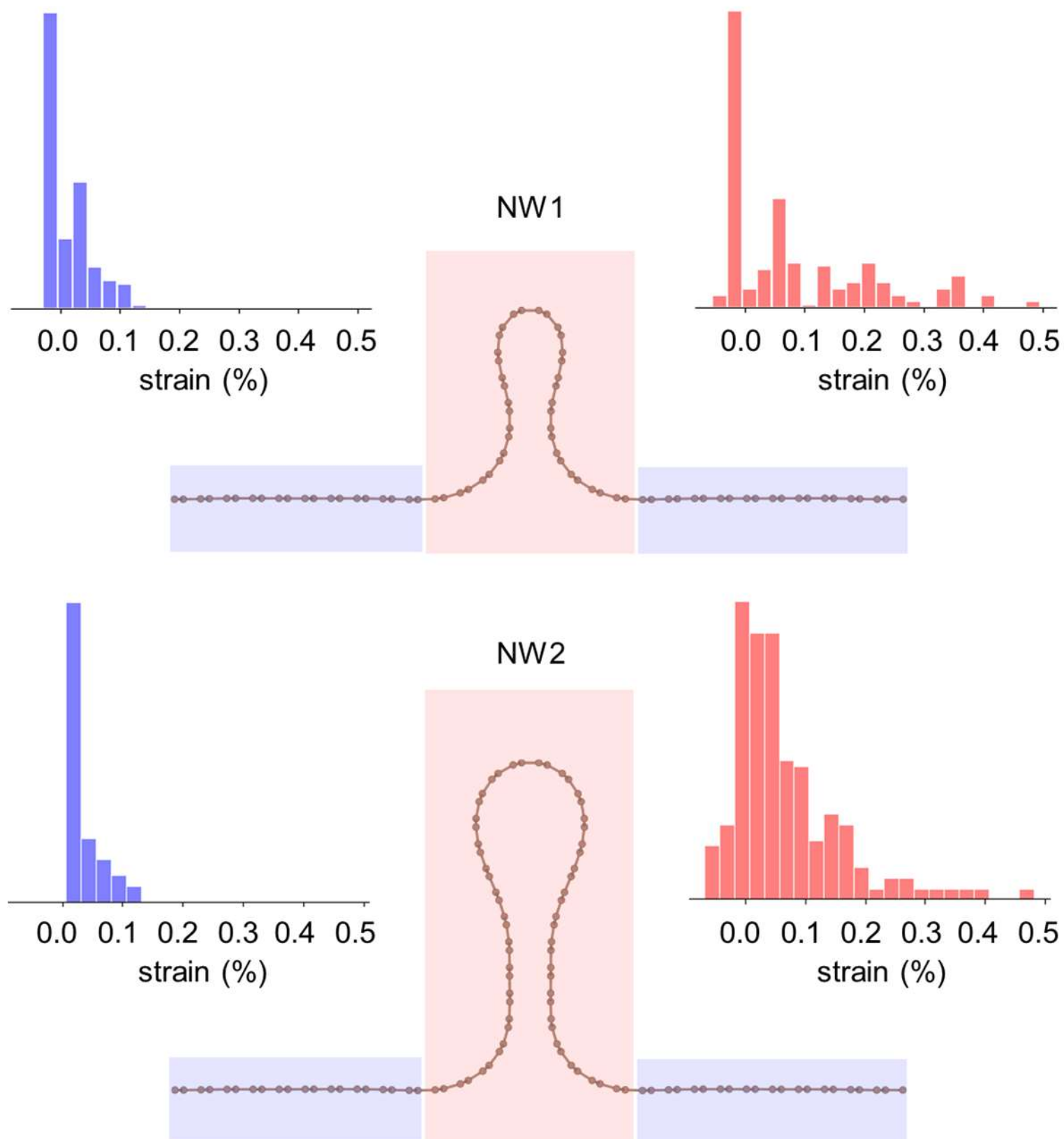

**Figure S6**: Strain lattice calculations for GNWr1 and GNWr2 (with continued GNWr3 in Fig. 1F).

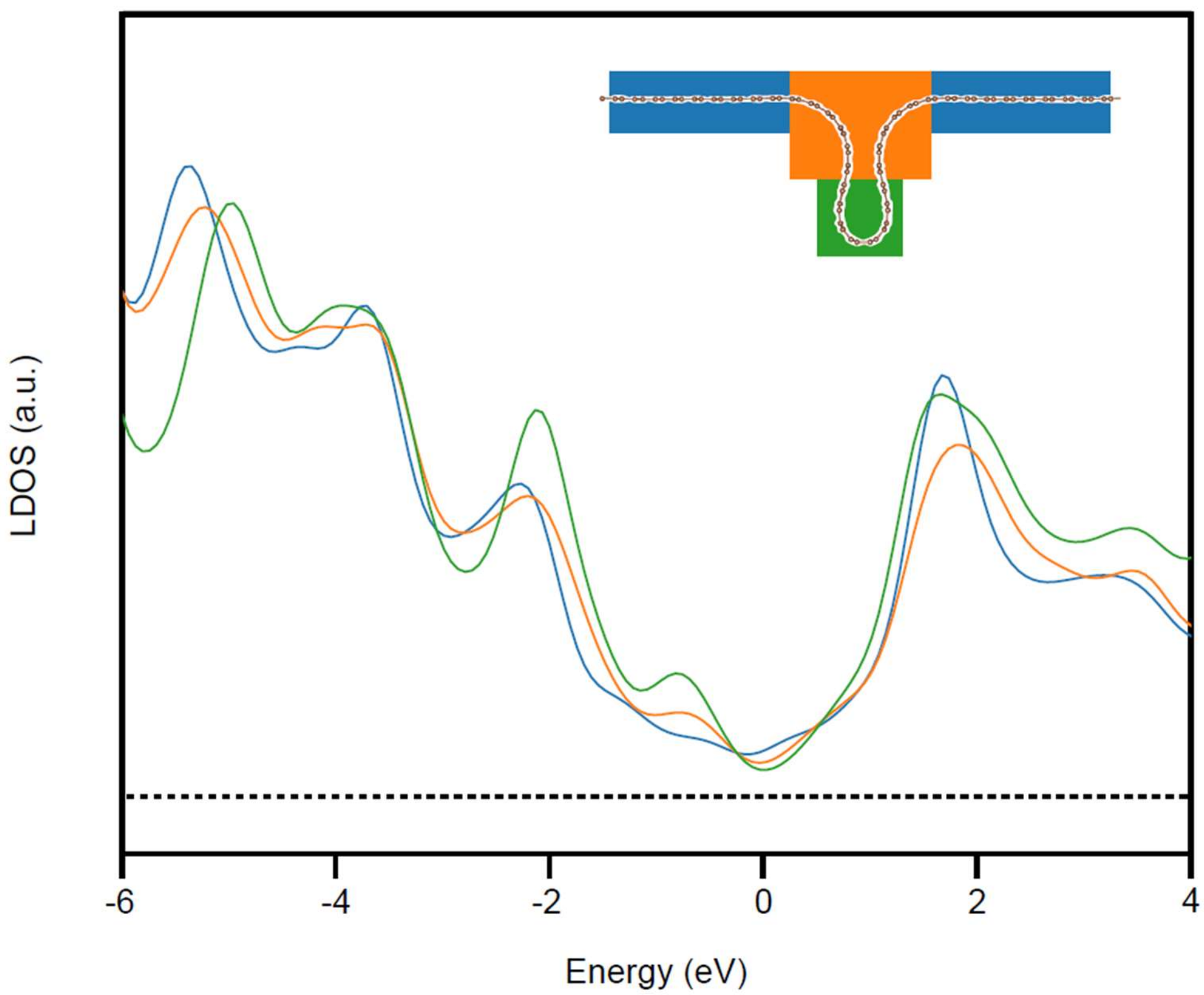

**Figure S7:** DFT local density of states (LDOS) averaged over three different areas of a GNWr, marked by the shaded areas in the insert.

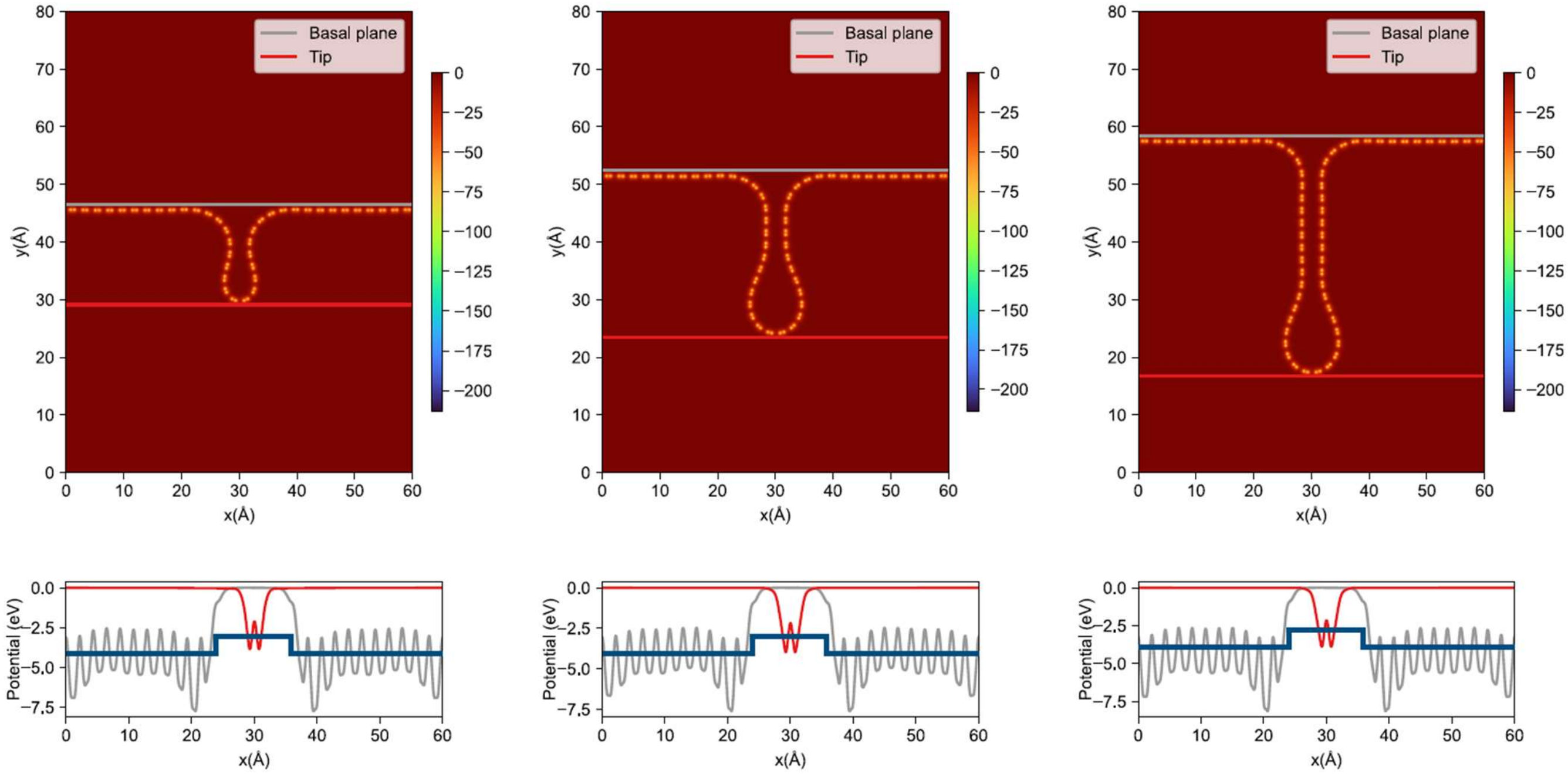

**Figure S8:** Potential in the absence of an electric field, clearly indicating the development of polarization due to the wrinkle– GNWr1, GNWr2, and GNWr3. In each case, this translates (bottom panel) to a potential barrier where electrons more easily escape the material due to the wrinkle. The computed band offset is about 1.2 V and does not significantly depend on the type of wrinkle.

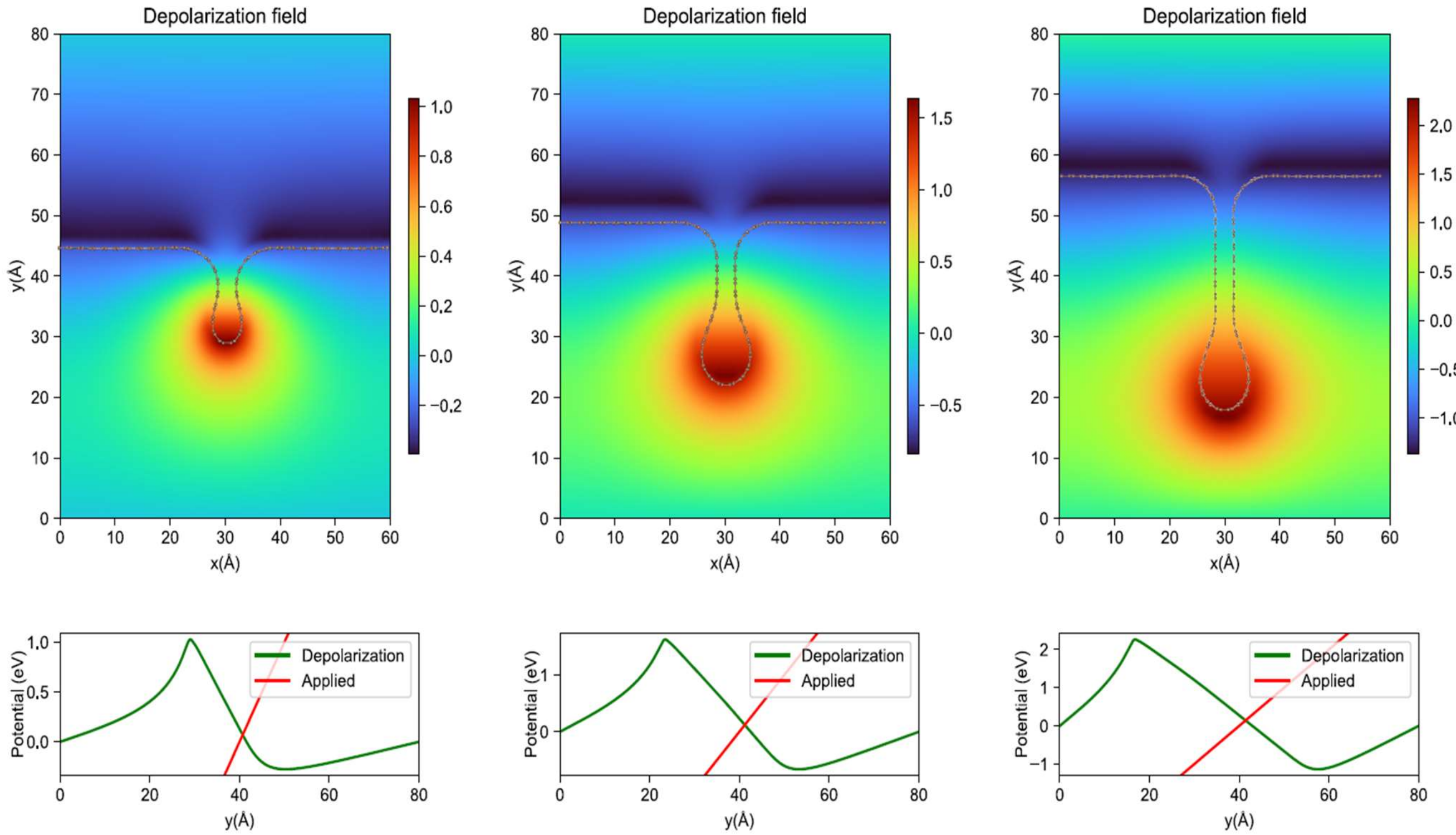

**Figure S9:** Depolarization field due to the presence of an external field shown in red in the bottom panel for GNWr1, GNWr2, and GNWr3.

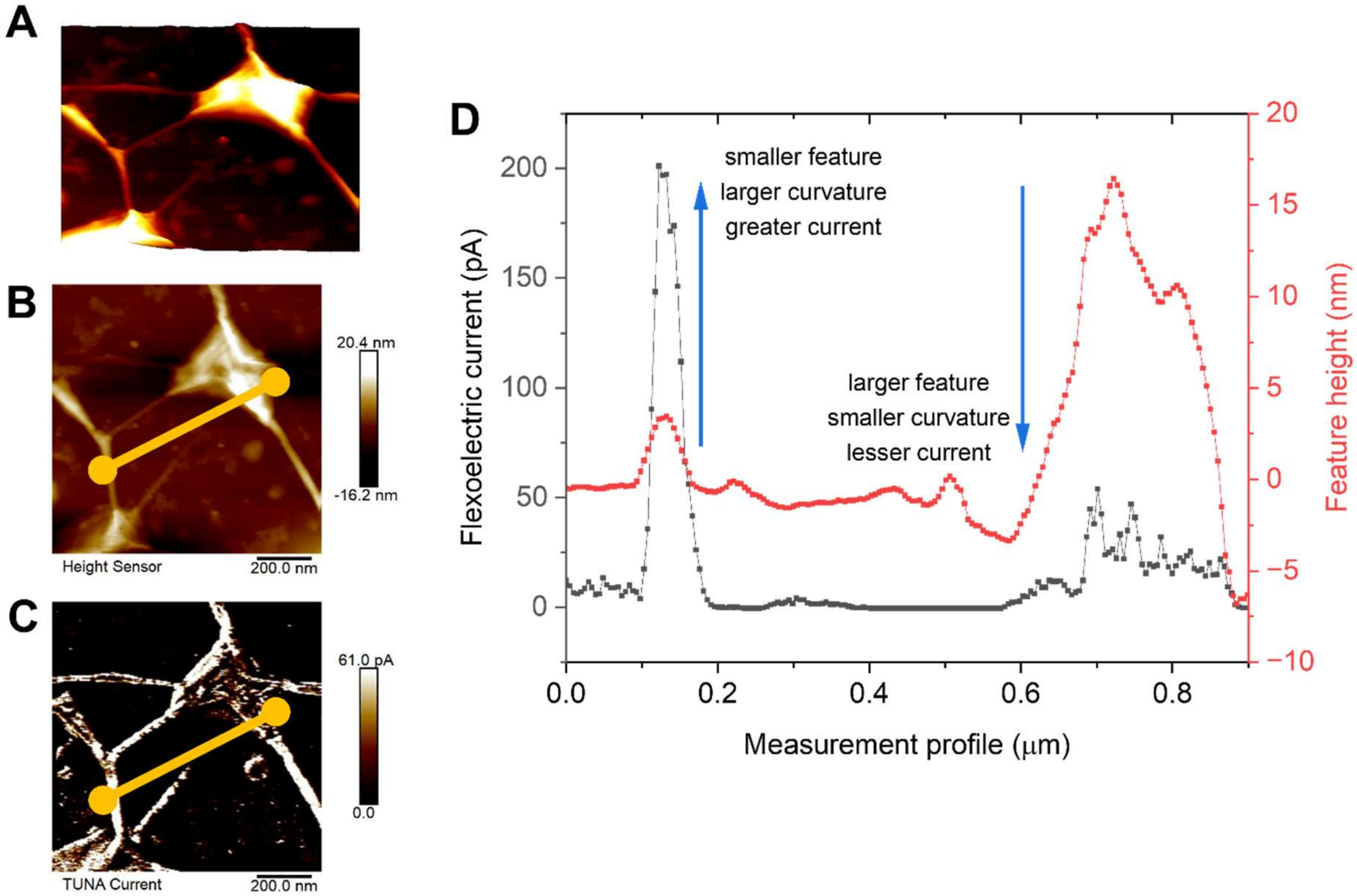

**Figure S10:** Direct comparison of the effect of curvature (and not height) on the flexoelectric current at a fixed potential. (A) 3D topology of features studied. (B) AFM micrograph. (C) C-AFM micrograph. (D) Comparing the topology and current across the drawn line profile. Note that features that do not have curvature but create variations in height (from 0.2 – 0.6 μm in the line profile along the x-axis), do not yield any current.

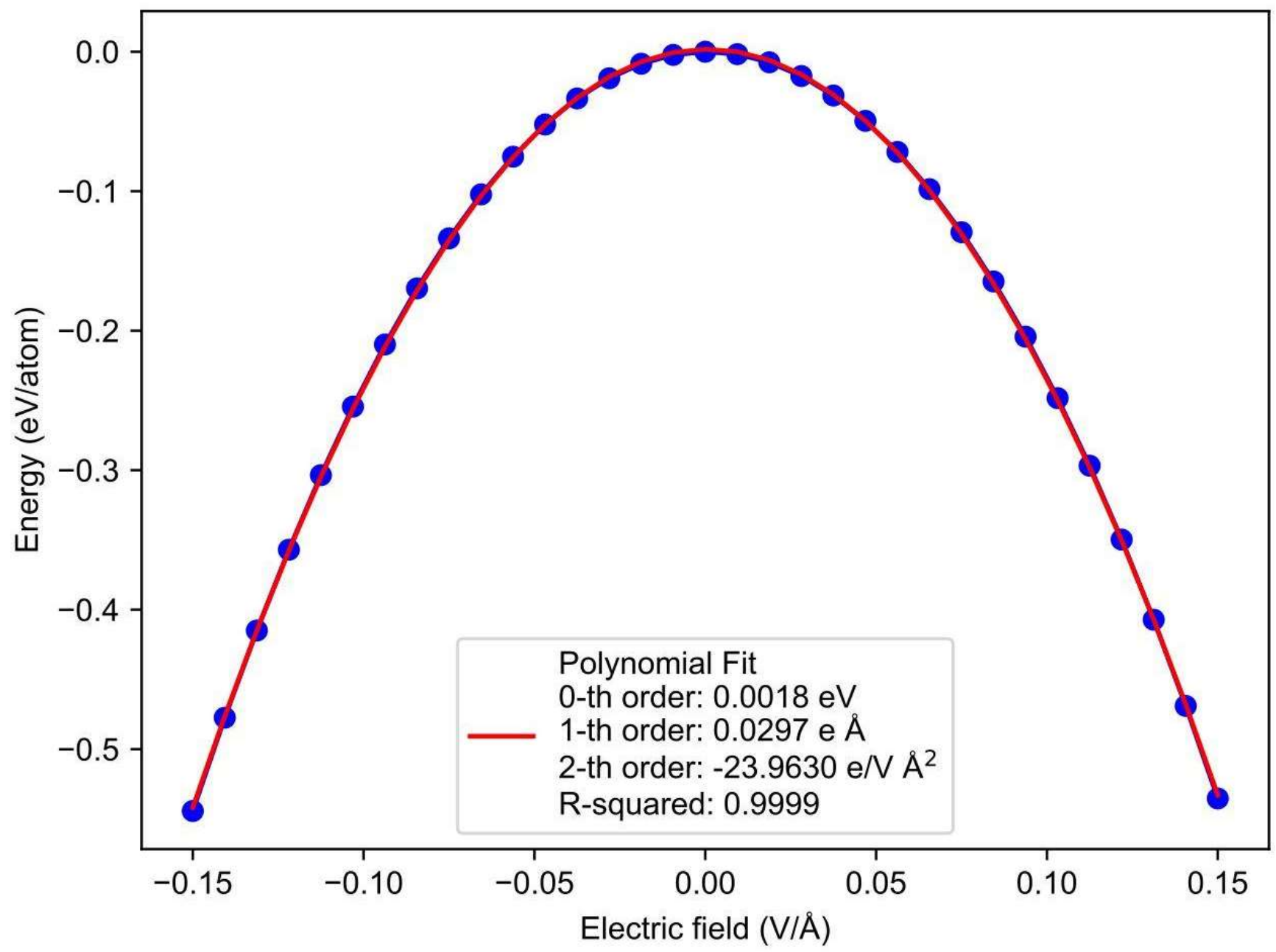

**Figure S11:** Energy as a function of applied field in the GNWr3 system shown in **Figure S3** (right structure). The induced dipole moment can be obtained from the fit to Equation s3.

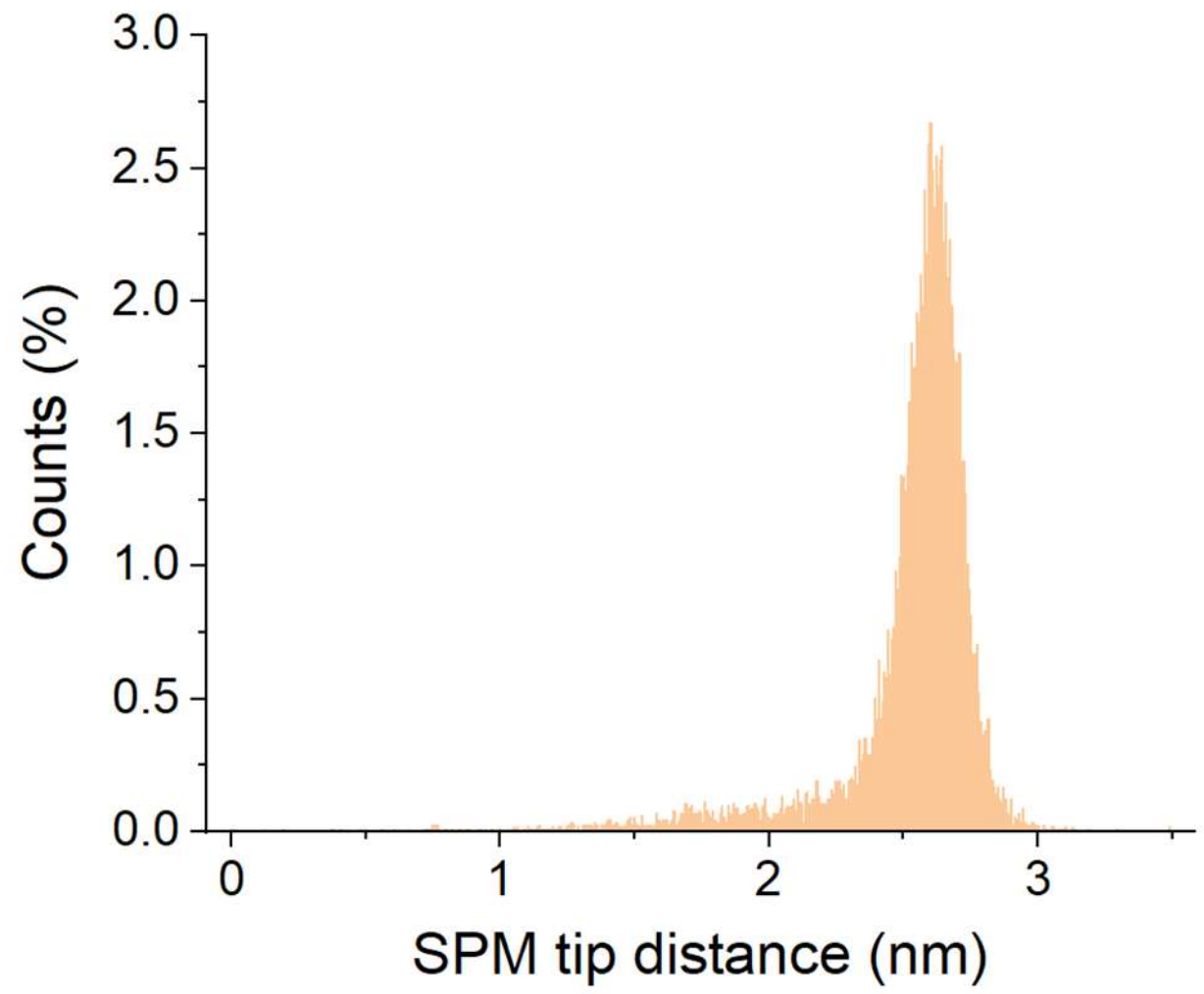

**Figure S12:** The histogram demonstrates the distribution of contact conditions, specifically the elastic deformation distance (in nanometers) associated with various wrinkles. The findings integrate the value of "jump to contact" and the minimal pressing distance to the surface, establishing a contact area between the probe apex and the GNWr. On a total of 512 pixels of data, the tip distance from the sample ($d_{spm\text{-}tip}$) is determined to be 2.52 ± 0.27 nm.